\documentclass[nofootinbib,showkeys]{revtex4}
\usepackage[utf8]{inputenc}
\usepackage{amsfonts}
\usepackage{amsmath}
\usepackage{amsthm}
\usepackage{amscd}
\usepackage{epsfig}
\usepackage{epstopdf}
\usepackage{graphicx}
  \graphicspath{{/}{figures/}}
  \DeclareGraphicsExtensions{.pdf,.png}
\usepackage{comment}
\usepackage{enumitem}
\usepackage{tabularx}
\newcounter{mycomment}

\newtheorem{Definition}{Definition}
\newtheorem{Theorem}{Theorem}

\begin{document}

%%%%%%%%%%%%%%%%%%%%%%%%%%%%%%%%%%%%%%%%%%%%%%%%%%%%%%%%%%%%%%%%%%%%%%%%%%%%%%%%%%%%%%%%%

\title{Anomaly on Superspace of Time Series Data}

%%%%%%%%%%%%%%%%%%%%%%%%%%%%%%%%%%%%%%%%%%%%%%%%%%%%%%%%%%%%%%%%%%%%%%%%%%%%%%%%%%%%%%%%%

\author{Salvatore Capozziello}
\email{capozziello@na.infn.it}
\affiliation{Dipartimento di Fisica, Universit\`a di Napoli ''Federico II'', Via Cinthia, I-80126, Napoli, Italy,}
\affiliation{Istituto Nazionale di Fisica Nucleare (INFN), Sez. di Napoli, Via Cinthia, Napoli, Italy,}
\affiliation{Gran Sasso Science Institute, Via F. Crispi 7,  I-67100, L'Aquila, Italy.}

\author{Richard Pincak}
\email{pincak@saske.sk}
\affiliation{Institute of Experimental Physics, Slovak Academy of Sciences, Watsonova 47, 043 53 Kosice, Slovak Republic,}
\affiliation{Bogoliubov Laboratory of Theoretical Physics, Joint Institute for Nuclear Research, 141980 Dubna, Moscow region, Russia.}

\author{Kabin Kanjamapornkul}
\email{kabinsky@hotmail.com}
\affiliation{Department of Survey Engineering, Faculty of Engineering, Chulalongkorn University, 254 Phyathai Road,
Bangkok, Thailand. }

\begin{abstract}
We apply the G-Theory and anomaly of ghost and anti-ghost fields in the theory of supersymmetry to study a superspace over time series data for the detection of hidden general supply and demand equilibrium in the financial market. We provide a proof of the existence of the general equilibrium point over 14-extradimensions of the new G-theory compared to M-theory of 11 dimensions model of  Edward Witten.  We found that the process of  coupling between nonequilibrium and equilibrium spinor fields of expectation ghost fields in the superspace of time series data induces an infinitely long exact sequence of cohomology from a short exact sequence of moduli state space model. If we assume that the financial market is separated into $2$ topological spaces of supply and demand as the D-brane and anti-D-brane model, then we can use a cohomology group to compute the stability of the market as a stable point of the general equilibrium of the interaction between D-branes of the market. We obtain the result that the general equilibrium will exist if and only if the 14-th-Batalin-Vilkovisky cohomology group with the negative dimensions underlying major 14 hidden factors influencing the market is zero.
\end{abstract}

%\pacs{11.15.Yc,  11.30.Pb, 87.14gn, 87.14gk}


\maketitle

\section{Introduction}\label{sec:intro}
%-------------------------------------------------------------------------------
The algebraic defect of mathematical modeling of the financial market over real number field is a source of extra dimensions in Kolmogorov space of underlying  time series data. We cannot observe hidden 14-extra dimensions in time series data because of all extra dimensions in 14-dimensional G-theory \cite{g,g2} of underlying time series data are canceled and they disappear in the analogy to the Feynmann model canceled half-way by the positive metric the longitudinal gauge particle, which is a real field. The source of the cancelation of the hidden dimensions is induced by the invariant property of the non-oriented supermanifold which is so-called parity  anomaly over supermanifold of ghost field \cite{anomaly}. An arbitrage is an evolution feedback between the hidden behavior of the trader from the ghost field of supply coupled with the anti-ghost field induced from the demand side and vanished when the market is in the general equilibrium. With this application of the cohomology theory in the economic as the quantum  fluctuation in the business cycle and the new G-theory, we produce a new market risk-cocycle coupled modeling over superspace in time series data. In  this new model, we can explicitly define an arbitrary opportunity as the Chern-Simon market anomaly over the induced ghost and anti-ghost field in the financial market.

Recently, most scientists \cite{stanley} and economists \cite{econ} turned their interest to use a new mathematical supergeometry \cite{super} and superalgebra \cite{super2} of Poisson bracket \cite{p}, an non-oriented supermanifold \cite{witten}, superpoint \cite{superpoint} and superstatistics \cite{super3,supersta} for the modeling of the anomalous equilibrium state \cite{stanley} in hidden dimensions of the financial market as the parallel worlds \cite{para} of the financial market in the general equilibrium theory. Typical economical modeling is based on the real number field in which each time series data of the record is embedded in the Euclidean plane without the supersymmetric property of the hidden ghost field as a supermanifold.

The supermanifold \cite{witten_int} is a high promising mathematical object in the mathematical physics which applies the mathematics of the Lie superalgebras to the behavior of bosons and fermions. The driving force in the formation of the supermanifold is a spinor field in mathematics and physics and the work on the characteristic class of the second cohomology group by Chern and Simon who also did a major contribution to use an anomaly current of the Yang-Mills field as the field with the hidden cancelation in the mirror symmetry modeling of the grand unify theory \cite{chern}. In the central concept of the supermathematics \cite{supermath}, the objects of study include the supersymmetry, supermanifolds,   the BV-cohomology \cite{bv} and the superlagrangian \cite{superla}, namely in the context of the superstring M-theory and G-theory \cite{g}. Besides using the dynamical stochastic general equilibrium model (DGSE)\cite{macro1,macro2} for the modeling of macroeconomics without taking into account the market microstructure, we can apply the supersymmetry theory from the superalgebra of the sheave sequence of the resolution theory from the BV-cohomology \cite{bv} and the unified theory of the anomaly group $SO(32)$ and the grand unified $E_{8}\times E_{8}$ \cite{e8} to unify the modeling of the microeconomics with the macroeconomics theory as the interaction of the D-brane and the anti-D-brane of IS-LM and DSGE in the supermanifold.

There exists a new approach \cite{pincak10} in the econophysics and the economic modeling \cite{handa} for the investigation of the general equilibrium modeling as a master equation for the orderbook in the stock market with the string \cite{pincak11,pincak12} and the D-brane theory approach \cite{pincak13}. The orderbook model based on the stochastic process \cite{orderbook} and defining a buy and sell operator as the time series model observes data directly, not from a hidden state. Most researchers who make a research on the financial market equilibrium \cite{pincak20} are coming from the signal processing, statistics, computer science and quantum physics area \cite{Sornette,Mantegna}, some of them are coming from the econometric research area studying the general equilibrium model \cite{econ} based on the dynamical stochastic general equilibrium model (DSGE) \cite{macro1,macro2} which is based on the stochastic control theory. We use an alternative approach in the cohomology \cite{bv} and the grand unified theory approach \cite{e8} to find a master equation for both microeconomics model lagrangian of the utility function throughout the macroeconomic model in the hidden superspace of the time series data. We called this equation, a new version of the modified Yang-Mill-Chern-Simon-master equation for the financial market over the ghost field in the financial market modeling.

In the space of time series data \cite{kolmogorov}, there is a source of hidden information while we record a data into the time series in a superspace, the so-called financial hidden ghost field. In the nature, all the physical quantities which have its pairs are so-called parity quantities. In quantum physics, the CPT-theory is an example of an active research beyond the superstatistics in which the theory is induced from the supermanifold. The generalisation of the supermanifold is the hypermanifold. Hyperstatistics is one of the examples of an active research for the M-theory and the unified theory with the hypersurface, namely Calabi-Yau space. In this work, we use this approach for the study of a new modeling of the quantum financial market in the supersymmetry theory of hidden negative 14 dimensions, the G-model.
In this work, we develop a new definition of the ghost field in the non-orientated state in the supermanifold modeling of the financial market. We introduce a new cohomology theory with a negative dimension with a unifying equation for the financial market in the superspace modeling.

The paper is organized as follows. In Section II, we specify the basic definition of the superspace in time series data and of the Lie superalgebra with the Poisson bracket in the master equation. We define 14-ghost fields in the financial market in this section. In Section III, we define a least action for the Yang-Mills field in the Chern-Simon theory in the financial market from the behavior of the traders in the financial market. We define a new BV-cohomology in the financial market with a new Yang-Mills-Chern-Simon-master equation for the financial market.
In this section, we provide a source of the proof of the existence of the general equilibrium model of the market in the grand unified theory, anomaly model in 14 hidden dimensions by using coupling between 14 ghost fields in the BV-cohomology theory.
  In Section IV, we discuss the result of the proof and we make a conclusion for future work.

\section{Superspace of time series data}
In the time series induced from the financial market, most researchers found their nonlinear and non-stationary state in which the normal distribution cannot be used while the market crashes. The source of the abnormal state in the financial time series might be a source of the hidden dimension in the Kolmogorov space of time series where ghost and anti-ghost fields of supply and demand of the financial market exists. Then we use the supersymmetry theory over the supermanifold of time series data. We can define a parity of the supply and demand by using the Poisson bracket. If we suppose that the supply $S$ is an inverse of the demand $D$ in the superspace of the financial market $S,D\in \mathcal{A}$, we can define a Poisson bracket $\{S,D\}$ in the financial market by using the adjoint and co-adjoint representations. We first introduce the classical master equation in the population based models in the standard probabilistic model. Then we extend the master equation to the superspace of time series data over the supermanifold with the induced ghost and anti-ghost fields in time series data.

\subsection{Chern-Simon (2+1) dimensions in 14 dimensional modelling of Financial Market}

In the macroeconomics, DGSE uses for the control of the inflation target $\pi_{t}$ of the superspace of macroeconomics as the financial market superspace in our model $\mathcal{A}$ the well-known instrument of the monetary policy interest rate, $i=i_{t}$. In the contrast to the microeconomics theory, there do not exist any inflation and interest rate in the utility and the production functions. Only the price and the quantity are defined there. Classical macroeconomic model is based on the stochastic model of the IS-LM model and the Phillips curve, so-called DGSE. The IS-LM model is based on the mathematical structure of a flat Euclidean plane without any curvature of the spacetime and with the mirror symmetry. We can visualize the IS-LM model (see Fig. \ref{fig1}) in the superspace of time series data by using the interaction between the D-brane and the anti-D-brane model, where the ghost field exists. One D-brane of the market is a supply side and the other anti-D-brane is the demand side of the market. With this model, equilibrium point in the IS-LM model will be connected by the components which imply that the supply side and the demand side are in each side of the M\"{o}bius strip of an non-oriented supermanifold (see Fig. \ref{fig2}).

\begin{figure}[!t]
\centering
\includegraphics[width=0.48\textwidth]{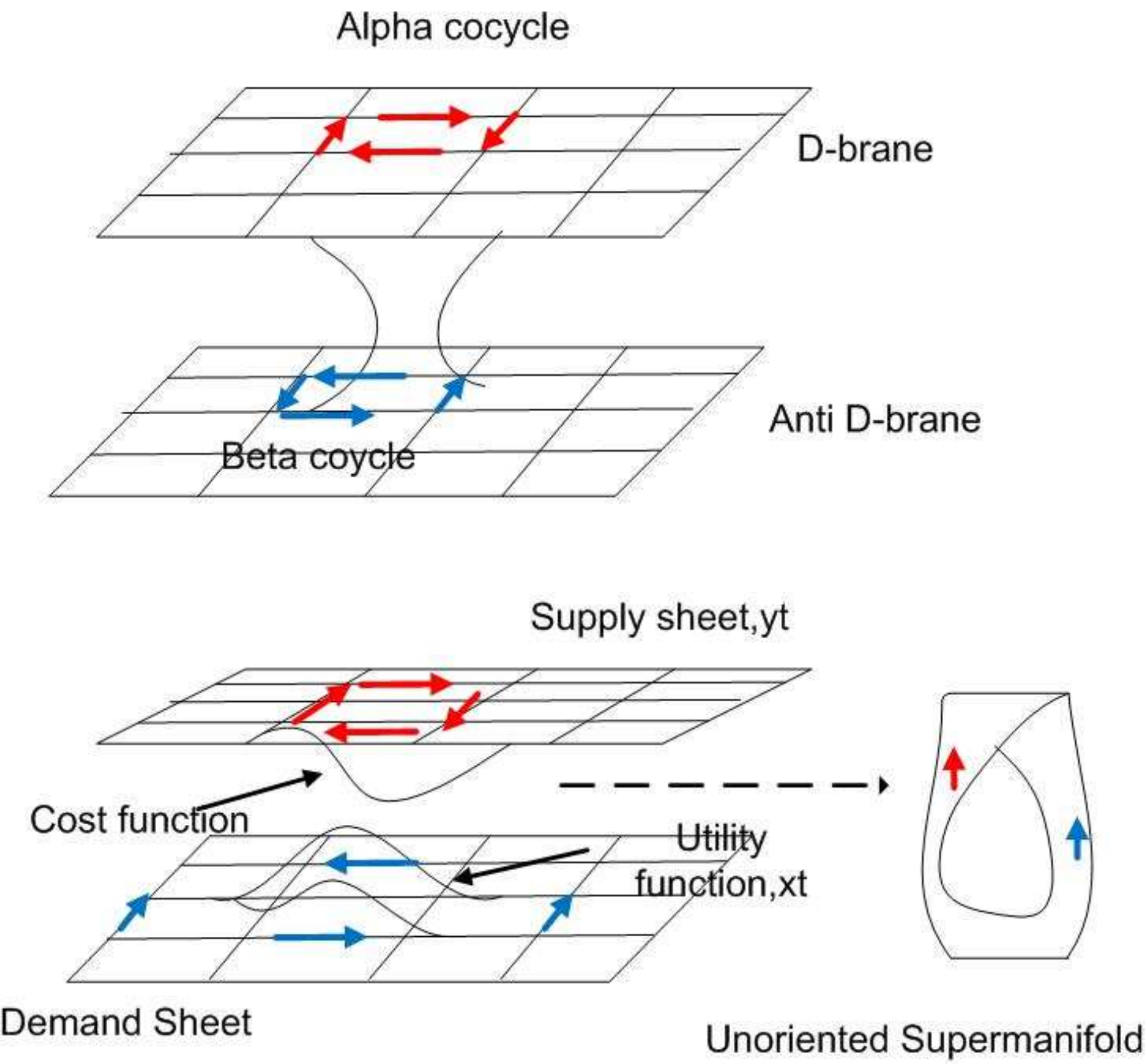}
\caption{Picture shows the market hypersurface of the supply side of as the D-brane and the demand side as the anti-D-brane. It is a superspace of time series data with the supersymmetry property of the mirror symmetry.
When the market is in the general equilibrium, we have an twist between the supply and the demand by using the duality property of the market as the duality map in the algebraic topology. This is the source of the non-oriented supermanifold in the financial market with the induce spinor field in time series data as the ghost and the  anti-ghost fields in the underlying extra dimension of the superspace in time series data.}
\label{fig5}
\end{figure}
In the history of the discovery of the ghost field, not many theoretical physicists notice the existence of ghost field from the formulas as a massless non-oriented ghost propagator in the Feynmann work on the loop space in the theoretical physics of the supersymmetry theory except Faddeev and Popov and D-Witt \cite{ghost} in the Yang-Mills theory. The hidden behavior of the traders on the expectation of the future market parameters is a source of the supersymmetry theory in the financial market where the ghost fields over the superspace of the market state in the financial market are defined over 14 hidden dimensions. There exists ghost and antighost fields in the financial market in which the hidden behavior of the traders is induced from the existence of the well-known market parameters (table  \ref{table1}). In the supersymmety theory, we have 2 mirror symmetries with an embedded non-oriented supermanifold between the 2 interaction planes of the D-brane and the anti-D-brane for the financial market (see Fig. \ref{fig5}).

The other source of the ghost field over the superspace of time series data is an algebraic defect of the real number field and the real vector space in the economic modeling. The superspace of the vector space over the real number field can be generalized by using the supermanifold model where the ghost field exists. Macroeconomics uses the Hamiltonian for the modeling of the macroeconomics model as the dynamics of the lagrangian of the market in the supply and the demand side.

In the theoretical physics, $U(1)$ gauge theory in space time coupled with the particle physics containing the massless charge $1$ is the most high promising theoretical structure for the grand unified theory or the M-theory for the  financial market modeling. In the Kolmogorov space in time series data, we can introduce a single massless Dirac fermion model for a new quantity (behavior of trader) as an extra dimension in the Chern-Simons term with $(2+1)$ dimensions \cite{simon} or what we called the Hopf term in the least action over the Yang-Mills field. In the theoretical physics and the string theory, we know that in the Chern-Simons term violate both the parity (P) and the time inversion (T) symmetries which are useful for the modeling of the gravitational field and the superstring theory. This problem gives rise to a new theory - so-called the parity anomaly on a non-oriented manifold. The application of this new theory now plays an important role in the topological superconductors, M-theory \cite{superpoint} and other applications to machine learning and the financial market modeling.

\subsection{Ghost Field in Financial Market}
Ghost field in the superspace of time series data can be visualized as the smallest unseparated component of time series data which we cannot separate and embed in the Kolmogorov space of time series data as a Calabi-Yau space (see Fig. \ref{fig3}).

Let $x_{t},y_{t}$ be 2 financial time series from the supply and the demand side. We can induce a field of a pair time series in the equilibrium state by a lag operator $g\in G$ of the Lie group translation

\begin{equation}
 Ad_{g}[x_{t},y_{t}]=[Ad_{g}x_{t},Ad_{g}y_{t}],
\end{equation}
where $Ad(g)=Ad_{g}.$ We have

\begin{equation}
ad=Ad_{\ast e}:\mathbf{g} \rightarrow End \mathbf{g}\nonumber,
\end{equation}

\begin{equation}
ad_{x_{t}}=\frac{d}{dt}|_{t=0} Ad_{e^{tx_{t}}}\nonumber,
\end{equation}

\begin{equation}
ad_{x_{t}}y_{t}=[x_{t},y_{t}].
\end{equation}

\begin{figure}[!t]
\centering
\includegraphics[width=0.48\textwidth]{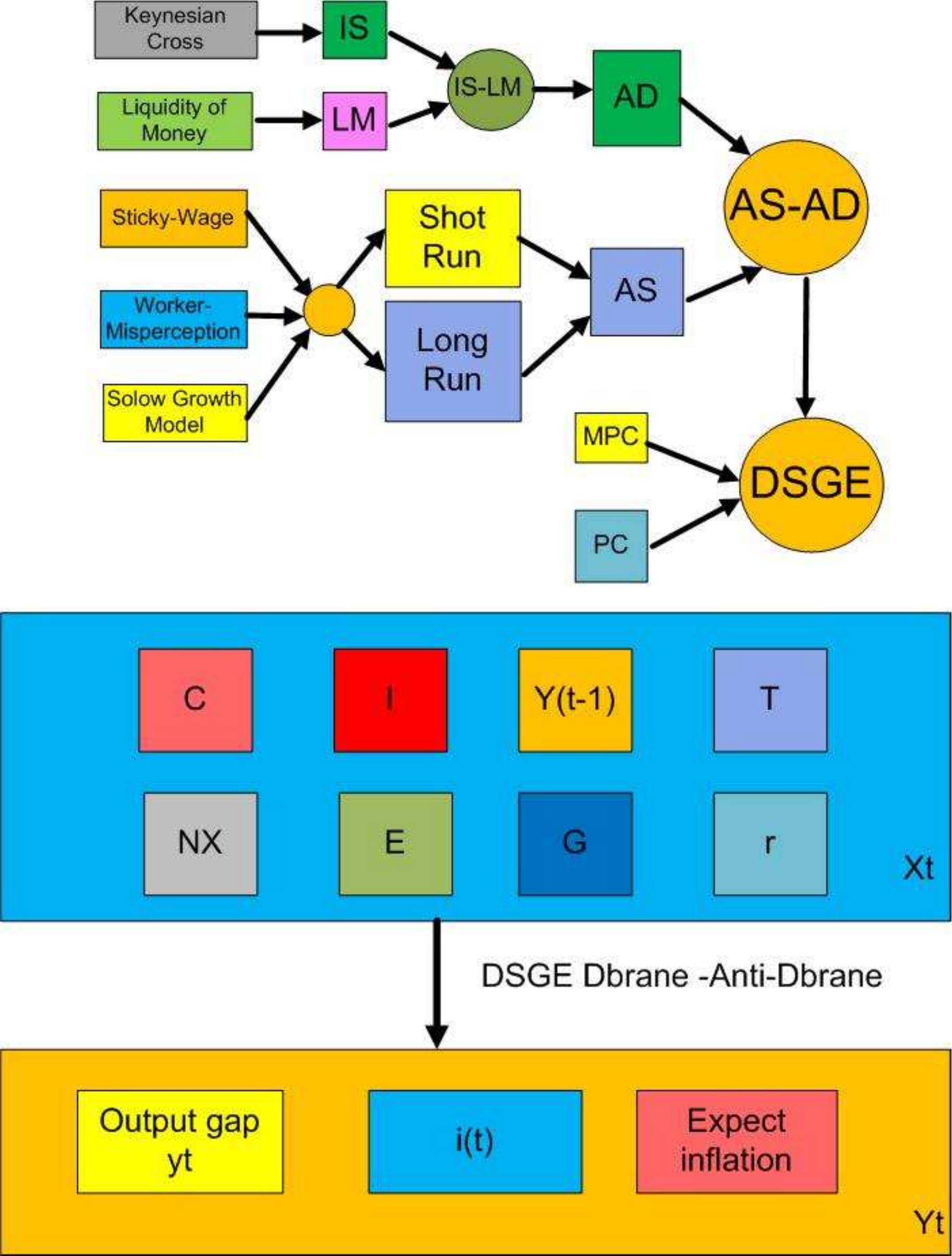}
\caption{Picture show the IS-LM and the DGSE model in the D-brane and the anti-D-brane diagram model. This is not the model with the same classical macroeconomics because we work on the level of the topological space underlying all the parameters in economics with an extra property of the spontaneous supersymmetry breaking. The equation of the IS-LM in the classical macroeconomic model is expressed by $Y=(Y-T)+I(r^{\ast})+G+NX(E).$ The meaning of the variable is shown in the table with its associated ghost and anti-ghost fields. The space of the demand side is a hidden space and denoted by $X_{t}$ and $Y_{t}$ is the supply side of economics in the IS-LM, AS-AD and DSGE model. All the demand and supply sides of economics have a hidden dimension of the underlying topological space defined by the supermanifold $(\mathcal{A},s)$ where the element of the supermanifold is a ghost or anti-ghost field of all the IS-LM parameters defined from the classical economics. This superspace of time series data of economics is defined by using the tangent space of the supermanifold of the underlying classical DSGE model.}\label{fig1}
\end{figure}

\begin{figure}[!t]
\centering
\includegraphics[width=0.48\textwidth]{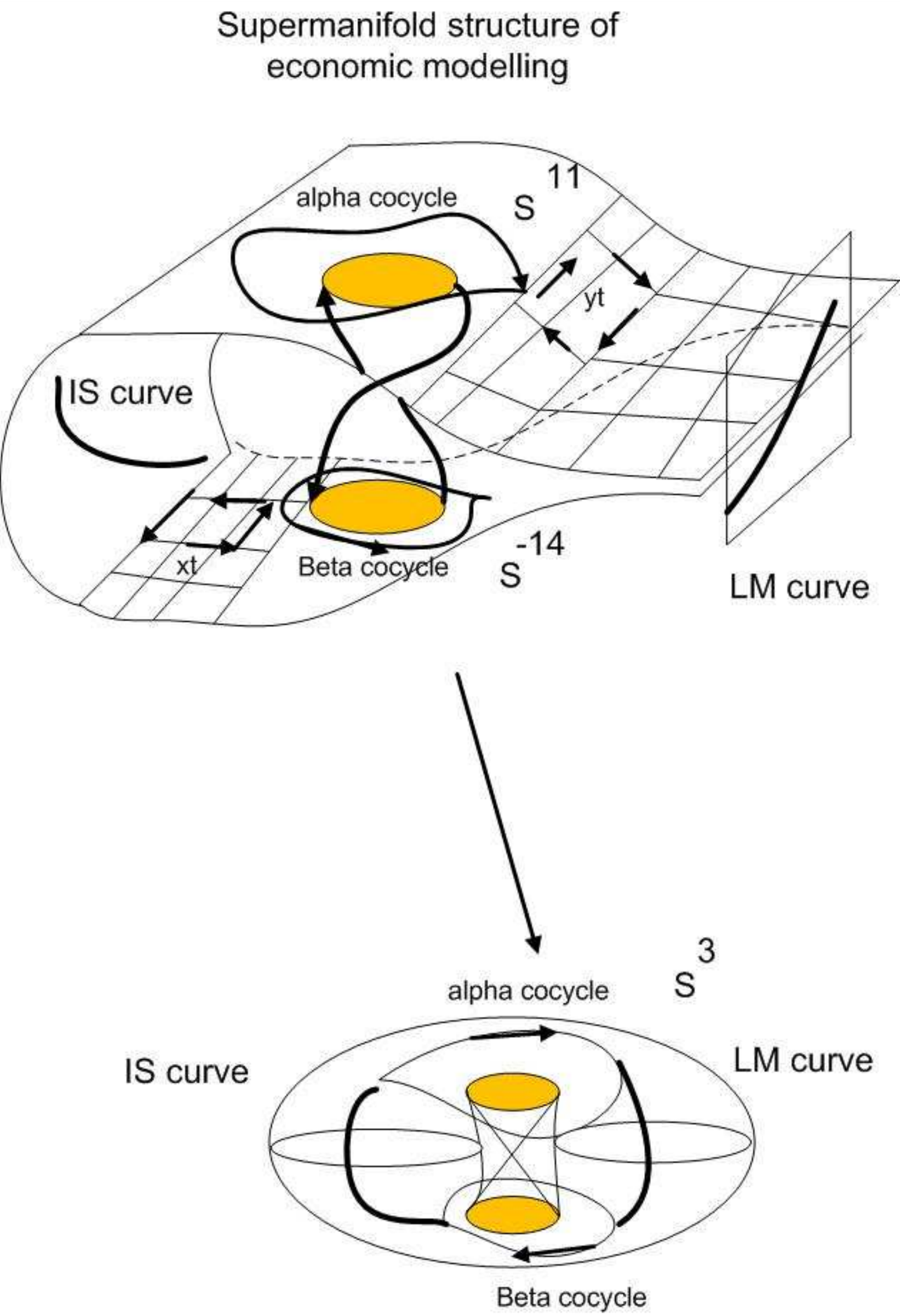}
\caption{Picture shows the four dimensional model with Hopf fibration $S^{3}$ of the IS-LM. We notice from the model of the superspace of time series data that in the IS curve a connected component with the LM curve and cannot be completely separated between the supply and demand side of the market. In the supply exists a demand and in the demand also exists a supply. When the market is in the equilibrium, the supply and the demand can be separated by using the market cocycles $\alpha_{t}$ and $\beta_{t}$ in the torus $S^{1}\times S^{1}$ as underlying space of the interaction between the D-brane and the anti-D-brane. $S^{-11}$ is defined as a hidden space with 11 factors influencing the macroeconomics on the supply side and $S^{-14}$ is a space with 14 dimensions influencing the market factor on the demand side. The details for these factors are shown in Table[xx] with its superspace of time series data in the ghost and anti-ghost fields.}\label{fig2}
\end{figure}

Let $\mathcal{A}^{j} $ be a superspace of time series data. The dimension of the superspace is denoted by the superscript $dim(A^{j})=j=gh(A)$ and can be plus and minus depending on the predefined ghost field in financial time series data.

\begin{Definition}
Every trader in the financial market has its ghost field to make a decision for buying and selling stock as  a decision field for sending command of the selling or buying message. We define two major types of this hidden decision field into 2 categories - the so-called ghost field $\Phi_{i}$ for buying operator (minimum state in the superspace of time series data) and the anti-ghost field $\Phi_{i}^{+}$ for selling operator (maximum state in the superspace of time series data).

\begin{equation}
\Phi: ( \mathcal{A},s) \rightarrow  \mathbb{Z}/2=\{1,-1\},
\end{equation}
where $\mathcal{A}$ is a supermanifold of the financial market with the master equation $s$ as a hidden ground field. Next, $[s_{4}]$ stands for the maximum state in the physiology of time series data as a ground field and $[s_{4}]$ stands for the minimum state as a ground field. The master equation is defined by the Poisson bracket of the superlagrangian of ghost and anti-ghost fields
\begin{equation}
 s:=\{ \int Sdt,-   \}=0
\end{equation}
where $S$ is the least action of the superlagrangian of the ghost field in the superspace of time series data.
\end{Definition}

Let $L$ be a lagrangian of the maximum utility function in the classical microeconomics model. We induce a superlagrangian $\mathcal{L}$ over the ghost field $\Phi(p)$ of price $p$. We assume that the ghost field is induced by the expectation price of the behavior of a trader in the financial market. If we are in general equilibrium of financial economics, we use the Batalin-Vilkovisky formalism (BV) for the doubling of the number of the fields in the financial market with parity (P). Let a price $p$ spinor field be denoted as $\Phi(p)$. Let $\Phi(p)^{+}$ be an antifield. The parity of the antifield $\Phi^{+}$ is according to the relationship $p(\Phi_{i}) =1-p(\Phi_{i}^{+})$ \cite{bv}, where $p(\Phi_{i})\in \{0,1\}\hspace{0.2cm} mod \hspace{0.2cm} 2$. Let $gh(\Phi_{i})$ be a ghost number of  $\Phi_{i}$. If $gh(\Phi_{i}(p_{t}))=0$, we come to the normal field of the price $p_{t}\sim N(0,1)$ in a stationary state in which it satisfies the random walk model. If $gh(\Phi_{i})>0$, we induce a ghost field of price. We can naturally assume that the market have 2 sides, the supply and the demand. In the superlagrangian model of the financial market, the ghost field will be on the supply side and the antighost field can be denoted as a hidden field on the demand side of the financial market in such a way that $gh(\Phi_{i})+gh(\Phi_{i})^{+}=-1.$ We use the BV cohomology for the financial market to formulate a Poisson structure between the ghost fields of the supply and the demand side of the market. We define a Poisson bracket of degree 1 by \cite{p}

   \begin{equation}
\{    \Phi_{i,S},\Phi_{j,D}^{+}\} =-\{  \Phi_{j,D}^{+},\Phi_{i,S} \} =\delta_{ij}.
 \end{equation}
Let $\Phi_{i,S} \in T_{g}G$ be a Lie superalgebra of the financial market and $\Phi_{j,D}\in  T_{g}G^{\ast}$ be a dual tangent supermanifold of financial market. We define the adjoint representation map for financial market by
\begin{equation}
ad_{\Phi_{i,S}}^{\ast}\Phi_{j,D}=\{\Phi_{i,S},\Phi_{j,D}^{+}\}.
\end{equation}
We have
\begin{equation}
ad_{\Phi_{i,S}}[\Phi(p),\Phi_{j,D}]=[ad_{\Phi_{i,S}}\Phi(p), ad_{\Phi_{i,S}} \Phi_{j,D}]
\end{equation}

\begin{figure}[!t]
\centering
\includegraphics[width=0.48\textwidth]{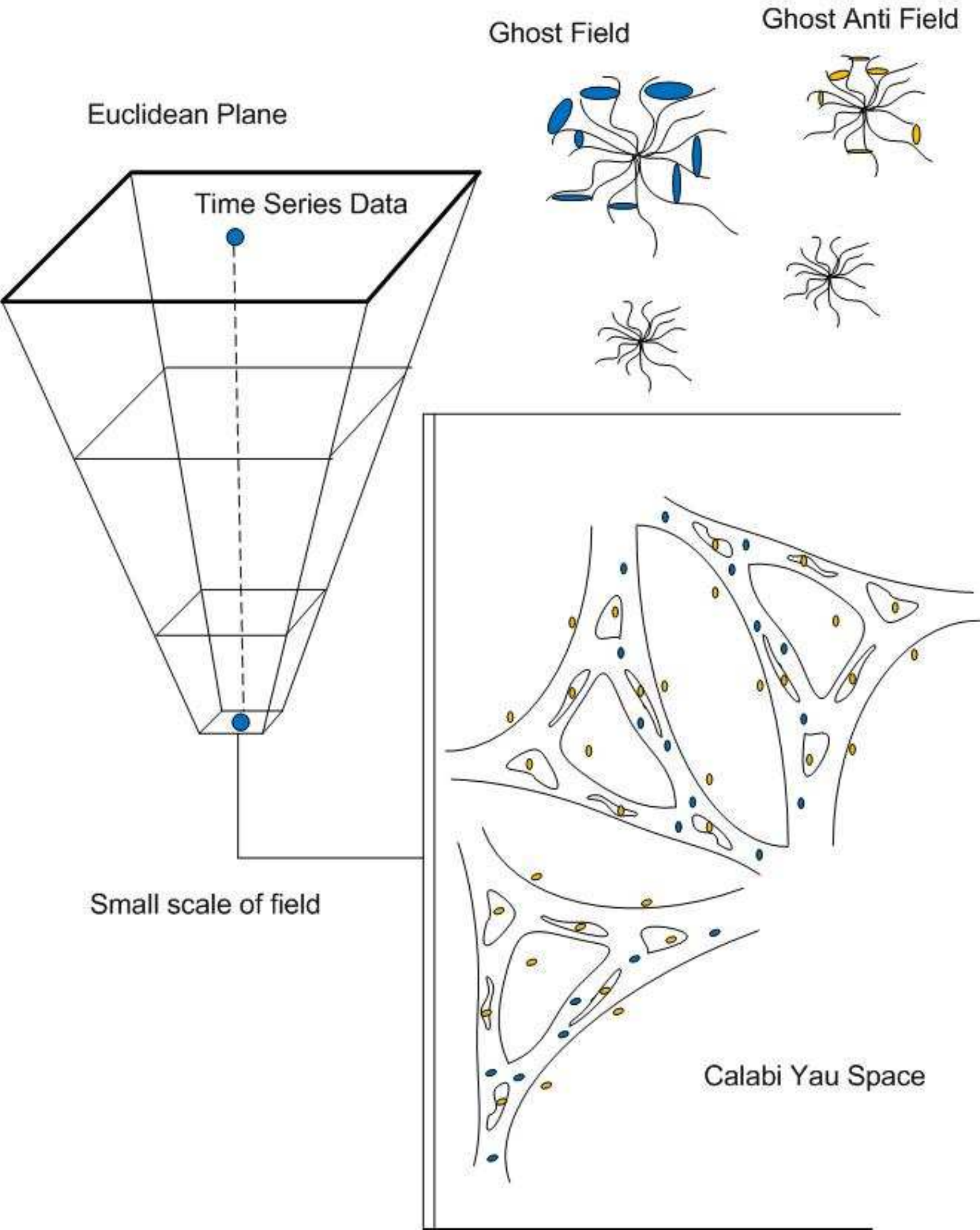}
\caption{Picture shows the Calabi-Yau space in time series data. It contains the ghost and anti-ghost fields in the superspace underlying financial time series data induced from the financial market. We cannot notice a Calabi-Yau space in time series data, because the scale of the Euclidean plane is to large to detect it. It curls into the extra dimension of 11-dimensions and 14-dimensions of underlying superspace of time series data. It is induced from the interaction of the ghost and anti-ghost fields of the behavior of the trader with the expectation fields of all the parameters in economics in which influences the market risk. If we work on the BV-cohomology level of the superspace of time series data, we can define the Calabi-Yau space in time series data embedded into the D-brane and anti-D-brane of the financial market microstructure.}
\label{fig3}
\end{figure}

We introduce 14 dimensional modelings of the financial market from both microeconomic and macroeconomic points of view with the anomaly gauge group in the $SO(32)$ G-theory. The source of 14 dimensions is defined as a hidden state for 12 dimensions and 2 states of the observed dimension in the spacetime of the pricetime of time series model. The definition of all the ghost fields in finance is the following:
\begin{itemize}
\item We use $\Phi_{i}(\pi),\Phi_{i}(r), gh(\Phi_{i}(\pi)+gh(\Phi_{i}(r))=-1$ for parity of ghost field for inflation and interest rate.
\item The source of more than six hidden dimensions induced from the ghost field of the Yang-Mills field in the financial market between the behavior of the traders in the market from the supply and the demand side and from the hidden supply and demand side. We denote $\Phi_{i}(F_{\mu\nu})+\Phi_{i}(\ast F^{\mu\nu})=-1 ,\mu=1,2,3 ,\nu=1,2,3.$
\item In the market, we have the supply ($S$) and demand ($D$) side, as the 2 dimensional model of the market is influenced by the market factors. We define their ghost field by
 $\Phi_{i}(S),\Phi_{i}(D), gh(\Phi_{i}(S)+gh(\Phi_{i}(D))=-1$.
\item The last 2 dimensions of the 14-dimensional model of the financial market is appear as time series data of buying and selling spinor operator in spinor field. We denote that ghost field by $(\theta^{\nu},x^{\nu})= (\Phi_{i}(B),\Phi_{i}(S)), gh(\Phi_{i}(B)+gh(\Phi_{i}(S))=-1.$
\end{itemize}
All predefined ghost fields carry their antighost field in the CPT-theory with parity $p(\Phi_{i})=1-p(\Phi_{i}^{+})$. We have a moduli state space model of ghost field and antighost field $mod\,2$ of its parity. We define a canonical coordinate for the Batalin-Vilkovisky antibracket by
\begin{equation}
                               \{   \Phi_{i},\Phi_{j}^{+}   \} =- \{   \Phi_{j}^{+},\Phi_{i}   \}=\delta_{ij}
\end{equation}

\begin{figure}[!t]
\centering
\includegraphics[width=0.48\textwidth]{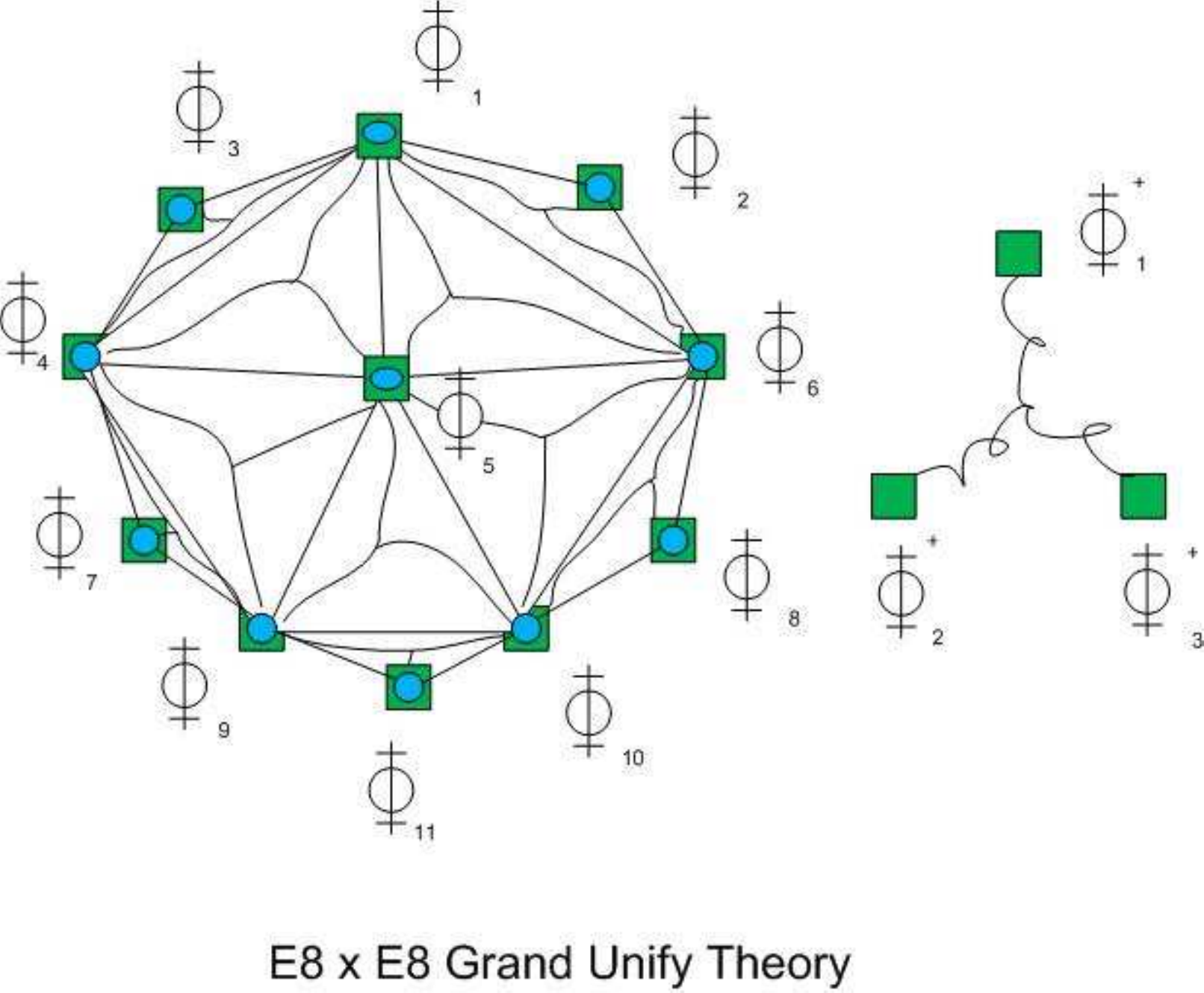}
\caption{Picture shows the icosahedral group $E_{8}\times E_{8}$ of the model from the grand unified theory for the financial market in the hidden 11-dimensional model of the supermanifold of the financial market as a D-brane with the ghost field of the cocycle $\beta_{t}\in H^{-14}(X_{t})$ quantum entanglement state with the anti-D-brane with $S^{11}$ with the induced ghost field of the cocycle $\alpha_{t} \in H^{11}(Y_{t}):=[Y_{t},S^{11}]$. If we consider the anomaly cancelation of the dimension from 14-11=3, we will get a state space model in spinor fields, $S^{3}$ as an error from the forecasting model in the econometric. In this way, we obtain a spinor field in time series data as a unified theory for the financial market. }
\label{fig6}
\end{figure}

Let $S$ be the least action of all the 14 ghost and anti-ghost fields in the financial market. For the first 6 ghost fields of the behavior of the trader in the market, we can write down an explicit form of $S$, a ghost field in the financial market by using a path integral of the least action borrowed from the formula for the gravitational Chern-Simons term in (2+1)-dimensional space time. It is one part of the grand unified theory in the theoretical physics (see Fig. \ref{fig6} for the details of 14 ghost fields in the G-theory). But in finance, the unified theory plays a very important role in the connection $\Gamma_{ij}^{k}$ of the Dirac spinor field from the theory of the supersymmetry.

\begin{table}[!t]

\renewcommand{\arraystretch}{1.3}

\caption{The table below shows the definition of the first 14 ghost fields in the financial market from the market microstructure of the underlying superspace in time series data. The ghost field is defined in extra dimensions of financial time series in which it is induce from the interaction of the behavior of the trader and the financial market. The underlying moduli state space $y_{t}$ in this example is defined as a macroeconomics parameter in the observation superspace. In fact, we can define $X_{t}$ be the space of the trade behavior and $Y_{t}$ be a space of the physiology superspace of time series data. But in this example, we apply it to the macroeconomic model as the superspace of the IS-LM and the DSGE model.}
\label{table1}
\centering
 \begin{tabularx}{\textwidth}{|c|c|X|c|c|c|} \hline \hline
 Ghost field $\Phi_{i}$  & parameter $(\beta_{i},\alpha_{i})$ &market hidden factors&market scale&state space &ground field \\ \hline \hline
    $\Phi_{1}$& $ U(x_{1},x_{2},\cdots )$& utility function &microeconomics & $x_{t}$&-1  \\ \hline

$\Phi_{2}$& $ I$& investment &macroeconomics&$x_{t}$
&-1 \\ \hline

$\Phi_{3}$& $G$&  government&macroeconomics&$x_{t}$&-1\\ \hline
$\Phi_{4}$& $ E$& exchange rate&macroeconomic&$x_{t}$&-1\\ \hline
$\Phi_{5}$&$r_{t}$ & real interest rate&macroeconomic&$x_{t}$&-1 \\\hline
$\Phi_{6}$ &  $i_{i}$ & MPC stimulus interest rate&macroeconomic&$x_{t}$&-1 \\\hline
$\Phi_{7}$& $\mathcal{D}$& demand side of market,utility function&microeconomic&$x_{t}$&-1\\ \hline
$\Phi_{8}$&      $T $ & Tax &macroeconomic &$x_{t}$  &-1\\ \hline
$\Phi_{9}$&   $  NX $& Net Export &  macroeconomic           &$x_{t}$&-1              \\ \hline
 $\Phi_{10}$& $A_{1},f_{-}$ & pessimistic fundamentalist trader&microeconomic &$x_{t}$&-1  \\ \hline
$\Phi_{11}$& $A_{2},\sigma_{-}$ & negative chat list trader&microeconomic  &$x_{t}$ &-1\\ \hline
$\Phi_{12}$&      $ A_{3},\beta_{-}$ & negative noise trader&microeconomic &$x_{t}$ &-1 \\ \hline
$\Phi_{13}$&   $ B$& buying operator &  microeconomic           &$x_{t}$          &-1     \\ \hline
 $\Phi_{14}$& $  M  $ & Money demand&macroeconomic &$x_{t}$ &-1 \\ \hline
$\Phi_{1}^{+}$& $\mathcal{S}$&supply side of market, lagrangian of production function&microeconomics&$y_{t}$&1\\ \hline
$\Phi_{2}^{+}$& $q=f(K,L)$& production function &microeconomics&$y_{t}$&1
 \\ \hline
$\Phi_{3}^{+}$&$A_{1},f_{+}$ & positive fundamentalist trader&microeconomics&$y_{t}$&1 \\\hline
$\Phi_{4}^{+}$ &  $A_{2},\sigma_{+}$ & positive chat list trader&microeconomics&$y_{t}$&1 \\\hline
$\Phi_{5}^{+}$&  $A_{3},\beta_{+}$ & positive noise trader&microeconomics &$y_{t}$&1 \\ \hline
$\Phi_{6}^{+}$&     $S$ & selling operator &microeconomics          &$y_{t}$  &1\\\hline
$\Phi_{7}^{+}$& $ Y$& output gap(DGP) &macroeconomics & $y_{t}$&1  \\ \hline
$\Phi_{8}^{+}$ &  $ \pi_{t}$ &  inflation&macroeconomics &$y_{t}^{\ast}$ &1\\ \hline
$\Phi_{9}^{+}$& $ U $ & unemployment rate  &  macroeconomics            &$y_{t}$ &1\\ \hline
$\Phi_{10}^{+}$    &    $p_{t}$ & price time series                       &micro/macro &$y_{t}^{\ast}$&1 \\\hline
$\Phi_{11}^{+}$ &     $[s_{i}]$ &physiology  of price time series                       &micro/macro   &$y_{t}^{\ast}$  &1\\ \hline \hline\hline
	 \end{tabularx}
\end{table}

%%%%%%%%%%%%%%%%%%%%%%%%%%%%%%%%%%%%%%%%%%%%%%%%%%%%%%%%%%%%%

\section{Anomaly in the Financial Market as an Arbitrary Opportunity}
\subsection{Yang-Mills Equation in Financial Market}
We study an arbitrary opportunity in the financial market by the induced hidden ghost field as the market cycle and cocycle $\alpha_{k},\beta_{k}\in H^{-14}(\mathcal{A},s).$ The 14 hidden ghost fields belong to the superalgebras of the supermanifold for the financial market $\mathcal{A}$ in which we cannot separated. A ground field $s$ of $H^{-14}(\mathcal{A},s)$ is a master equation of the transition superprobability of coupling 2 market cocycles as a spinor field in financial time series data (see Fig. \ref{fig4}). The way in which the cocycles couple each other in the hidden dimension, we cannot observe in space of time series data. The coupling induces a market quantum non-conserved chiral current $J^{\mu}_{A}$ and varies from an arbitrary opportunity in the financial market in which it is defined from the behavior of the trader in the Eric transactional model of the behavior of the trader in the new cohomology theory in the financial market\cite{cohomo}.

We define a Yang-Mills field for the financial market by the interaction of the Chern-Simon-Eric field
 $\mathcal{A}$ induced by the interaction of 2 sides of the market of the behavior of the traders $A_{i},i=1,2,3$.

We are going to prove the existence of the tensor field $g_{ij}$ in $\alpha_{t}$ and the tensor field $g^{ij}$ in   $\beta_{t}$ as an interaction between 2 D-branes of an induced third tensor field in orbifold $g_{i}^{k}$ as an ingredient of the connection in the financial market.

Consider the interaction between the ghost and anti-ghost fields in the financial market.

\begin{figure}[!t]
\centering
\includegraphics[width=0.48\textwidth]{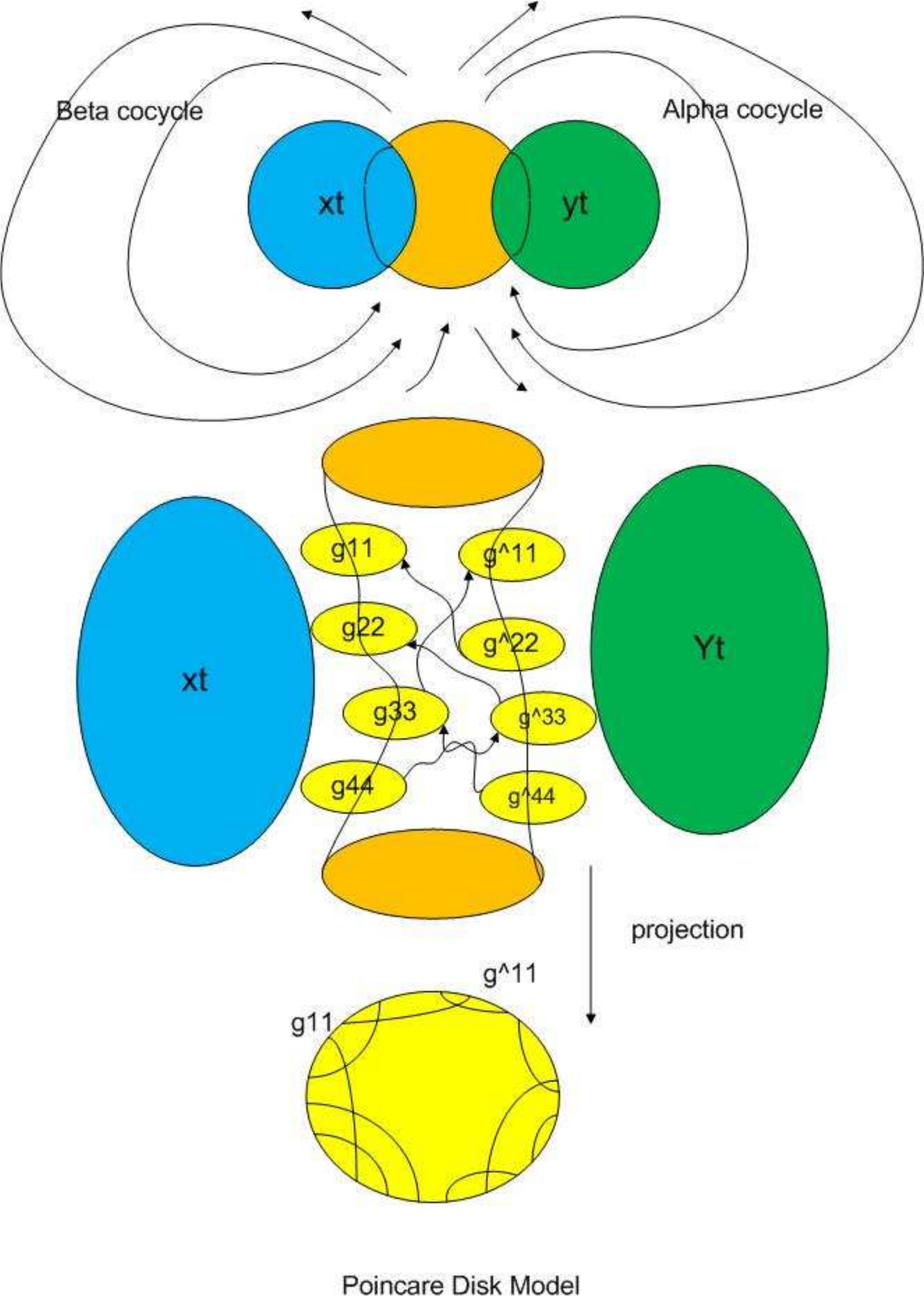}
\caption{Picture shows the interaction between the berazian coordinate in the supermanifold of time series data $X_{t}$ and $Y_{t}$. We simplify its berazian by borrowing a notation of the jacobian field as the new definition of the tensor ghost field in financial time series data. An arbitrary opportunity is directly defined by changing the tensor ghost field using the interaction of the behavior of the trader in the market. We can use the Poincare disk model to visualize the market equilibrium by denoting one circle of the supplied field. The other induced cycle is the demand field. If these two cycles contact to each other properly, the market will be in equilibrium. If the circle is not in the proper position, all field will be induced and push each other as the market cocycle model.}
\label{fig7}
\end{figure}
\begin{Definition}
The tensor field of the berazian coordinate transformation in the financial market is defined by the change of the state of the ghost and anti-ghost fields in finance (see Fig. \ref{fig7}):
\begin{equation}
g_{ij}=\frac{\partial \Phi_{i}}{\partial \Phi_{j}}, \hspace{0.5cm}
 g^{ij}=\frac{\partial \Phi_{i}^{+}}{\partial \Phi_{j}^{+}}
 ,\hspace{0.5cm}g_{j}^{i}=\frac{\partial \Phi_{i}}{\partial \Phi_{j}^{+}}.
\end{equation}
\end{Definition}

\begin{Definition}
The connection over the ghost field in the financial market is defined by
\begin{equation}
  \Gamma_{ij}^{m}=\frac{1}{2}g^{ml}(\partial_{j}g_{il} +\partial_{i}g_{lj}-\partial_{l}g_{ji}).
\end{equation}
\end{Definition}

\begin{Definition}
The Ricci tensor in time series data is a contraction of the curvature tensor defined by
$R_{ik}=R_{ikl}^{j}$ with respect to the natural frame of an arbitrary opportunity (connection),
\begin{equation}
R_{ik}=\partial_{k}\Gamma_{ji}^{j}- \partial_{j}\Gamma_{ki}^{j}+\Gamma_{km}^{j}\Gamma_{ji}^{m}-
\Gamma_{jm}^{j}\Gamma_{ki}^{m}.
\end{equation}

\end{Definition}

\begin{Definition}
Let $F_{\mu\nu}$ be a Yang-Mills field for the financial market and let $A_{\mu}$ be a Chern-Simon field for the financial market. Let $H^{2}_{DR}(\mathcal{A})$ be a De-Rham cohomology for the financial market $\mathcal{A}$. We have a supersymmetry of anti-self dual between the brane and the anti-brane in the financial market defined by

\begin{equation}
\ast F^{\bigtriangledown_{[s_{i}]}}
=\oint _{H^{2}_{DR} (\mathcal{A}) }  F_{\mu\nu}^{k} A_{\nu} =F^{\bigtriangledown_{[s_{i}^{\ast}]}}
\end{equation}

\end{Definition}

\begin{Definition}
We have the {\em anomaly} or the arbitrary opportunity after the quantization of the market by the behavior of the trader $F^{\mu\nu}$ with the anti-D-sister  $F^{\bigtriangledown_{[s_{i}]}}$ with the Chern-Simon current induced in hidden -14 extra dimensions of financial market:
\begin{equation}
 J_{A}^{\mu}=\partial \mu F^{\mu\nu} +[A_{\mu},F^{\mu\nu}].
\end{equation}

\end{Definition}

\begin{equation}
\frac{1}{4}<\ast F^{\mu\nu} F_{\mu\nu}>=\partial K^{\mu},
\end{equation}
where
\begin{equation}
K^{\mu}=\epsilon^{\mu\alpha\beta\gamma} <\frac{1}{2} A_{\alpha} \partial_{\beta} A_{\gamma} +\frac{1}{3}
A_{\alpha} A_{\beta} A_{\gamma}>,
\end{equation}
where a connection $\Gamma_{\alpha\nu}^{\mu}=(A_{\alpha})_{\nu}^{\mu}$, where $A$ is a Chern-Simon term in the financial market. An arbitrage or arbitron is a Chern-Simons anomaly, $K^{\mu} $ or an anomaly current induced from the behavior of the trader as the twist ghost field between 2 sides of the market. This field of the financial market induces a Ricci curvature in the financial market as an arbitrary opportunity for each connection in the coupling ghost field of the behavior of the trader. This curvature blends the Euclidean plane of the space of the observation to twist with the mirror plane behind the Euclidean plane and wraps to each other as the D-brane and the anti-D-brane interaction of the superspace of financial time series data (see Fig. \ref{fig1}).

\begin{equation}
R^{\mu}_{\nu\alpha\beta}(\Gamma) = e_{\alpha}^{\mu} R_{b\alpha\beta}^{a} e_{\nu}^{b}.
\end{equation}

\begin{Definition}
Let $\mathcal{C}$ be a configuration space of the subsystem of the financial market and denote the probability of finding a system  in the configuration of the supply side $\mathcal{C}$ at time $t$. Let $\mathcal{C}'$ be a configuration space of the subsystem and denote the probability of finding a system in the configuration of the demand side of the financial market $\mathcal{D}$ at time $t$ by $\mathcal{P}_{\mathcal{D}(t)}$. A master equation is a dynamics of the financial market system of the transition probability between the supply and the demand state defined by
\begin{equation}
\frac{d\mathcal{P}_{\mathcal{D}(t)}}{dt}=\sum_{\mathcal{S}_{i}}\lambda_{\mathcal{S}\mathcal{D}} \mathcal{P}_{\mathcal{S}(t)}-\sum_{\mathcal{S}_{i}}\lambda_{\mathcal{D}\mathcal{S}} \mathcal{P}_{\mathcal{D}(t)}
\end{equation}
where $\lambda_{\mathcal{S}\mathcal{D}}$ is a transition probability between the supply and demand market state in the financial market.
\end{Definition}

\begin{Definition}
The classical master equation in the Batalin-Vilkovisky cohomology for the financial market is defined by the solution of the classical master
\begin{equation}
\{ \int \mathcal{P}_{\mathcal{S}}(t)dt,\int \mathcal{P}_{\mathcal{D}}(t)dt\}=:\{ \Phi_{i}(x_{t}),\Phi_{j}^{+}(y_{t})\}=  0,
\end{equation}

where $\mathcal{P}_{\mathcal{S}}(t)\in \mathcal{A}^{0,0}$ is a differential expression in the ghost and anti-ghost fields with $gh(S)=0$, a ghost number and a parity number $ p(S)=0$. Here, $x_{t}$ is the demand side of the market and $y_{t}$ is the supply side of the market.
\end{Definition}

\subsection{Anomaly on a non-oriented supermanifold of financial market}

A coordinate of the space of time series is invariant under the group transformation - we call it the translational and rotational invariant, but not for the reflection of the time scaled (R) and the time reversal (T) symmetry.

Let $x_{t}\in X$ with the dimension of $X$ in $(2+1)$ dimension, we quantize the gauge Chern-Simon field in time series data $x_{t}$ by the least action as a ghost field $S$.

Let $\mathcal{A}$ be a superspace, where the financial market of the measurement is induced by its ghost field. Let $f\in \mathcal{A}\in \mathcal{F}$, we have a probabilistic density in the superstatistics $p(x_{t})=\int f dt$ in $\Phi_{i}$. Let $x_{t}$ be a financial time series of the observation in moduli state space model of the supermanifold. Let $G$ be the ghost field of buying and selling operators. Let $S$ be the least action of the financial market as the ghost field. We have a relationship
\begin{equation}
 \{  \{G,S\}  , x_{t}\} =\partial x_{t}.
\end{equation}
The BV cohomology for the financial market is defined by the differential with the exact sequence with $s^{2}=0$.
We define a canonical coordinate of the financial market in the ghost field of 14 dimensions by using the Poisson bracket as the codifferential complex in the BV-cohomology of the master equation in the financial market,
\begin{equation}
s=\{ \int S dt,-\}.
\end{equation}
We induce a BV-cohomology as long exact sequence of the financial market as followings,
\begin{equation}
 \cdots \rightarrow H^{-11}(\mathcal{A},s)\rightarrow \cdots   \cdots \rightarrow H^{-1}(\mathcal{A},s)
\stackrel{\partial}{\rightarrow} H^{-1}(\mathcal{A},s)\rightarrow H^{-1}(\mathcal{\mathcal{F}},s)\rightarrow H^{0}(\mathcal{A}/C,s)\rightarrow \nonumber
\end{equation}

\begin{equation}
 \rightarrow H^{0}(\mathcal{A},s) \rightarrow H^{0}(\mathcal{F},s)\rightarrow H^{1}(\mathcal{\mathcal{A}},s)
\stackrel{\partial}{\rightarrow} H^{1}(\mathcal{F},s)\rightarrow  \cdots \rightarrow H^{3}(\mathcal{\mathcal{A}},s)\rightarrow \cdots
\end{equation}

\begin{Definition}
We define a differential of the ghost and anti-ghost fields with parity by
\begin{equation}
s\Phi  =(-1)^{p(\Phi_{i}) +1}\frac{\delta S}{\delta \Phi_{i}^{+}},
\end{equation}

\begin{equation}
s\Phi^{+}  =(-1)^{p(\Phi_{i})}\frac{\delta S}{\delta \Phi_{i}}.
\end{equation}
\end{Definition}

\begin{figure}[!t]
\centering
\includegraphics[width=0.48\textwidth]{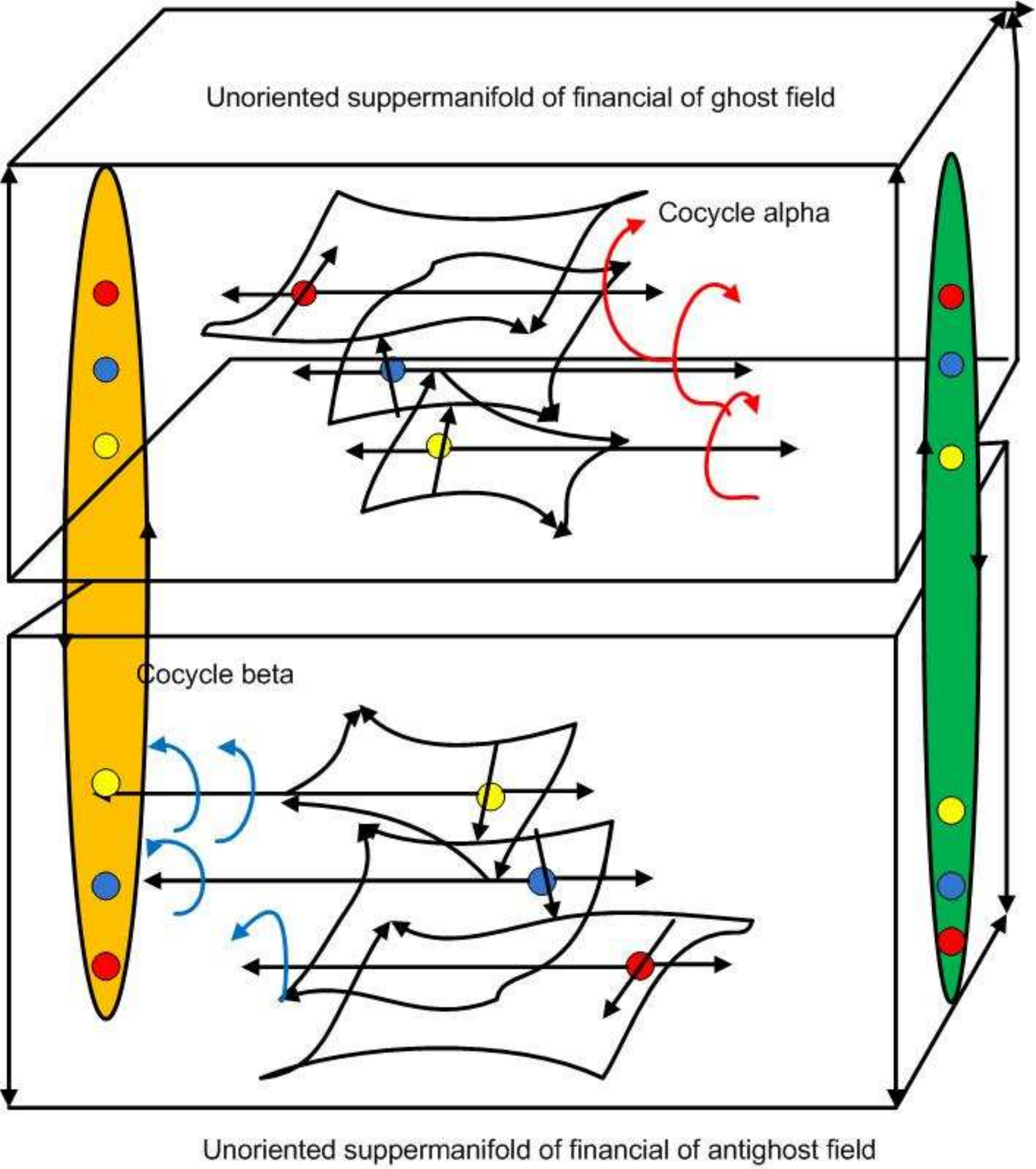}
\caption{Picture shows spin-orbit coupling between the ghost and anti-ghost field cocycles and localized in the reverse direction of the time scale. The down box is the anti-D-brane of the financial market and the upper box is the D-brane of the financial market. We cannot notice the arbitron like the arbitrary opportunity particle, because the collision and the dynamics of the superprobability transition is very fast and exists at a very small scale, so-called Calabi-Yau space of time series data.}
\label{fig4}
\end{figure}

\subsection{Superstatistics of superpoint and ordinary least square regression(OLS)}

We start from the normal situation of the normal distribution of the shock in the economics $\epsilon \sim N(\mu,\sigma^{2})$ in the ordinary statistics in the sigma field,
\begin{equation}
y_{t}=\alpha_{t}+\beta_{t}x_{t} +\epsilon_{t},
\end{equation}
where we have a new integration over the cocycle $\alpha_{t}$ and $\beta_{t}$ in the BV-cohomology.

This is a new kind of a supermathematical theory in the probability theory with a new integral sign over the tangent of the supermanifold under the Berezin coordinate transformation. The structure of the supergeometry of the superpoint \cite{superpoint} induces a superstatistics of the hidden ghost field in the financial market. We let $y_{t}$ be an observable variable in the state space model and $x_{t}$ be ahidden variable of the state. We denote $\Phi_{i}(y_{t})\in \mathcal{A}$ be a ghost field and $\Phi_{i}^{+}(x_{t})$ be an anti-ghost field in finance with the relation of its ghost number,

\begin{equation}
gh(\Phi_{i}(y_{t}))+gh(\Phi_{i}^{+}(x_{t}))
=-1.
\end{equation}
We have a parity $p(\Phi_{i}(y_{t}))=1-p(\Phi_{i}^{+}(y_{t})) $ and $p(\Phi_{i}(x_{t}))=1-p(\Phi_{i}^{+}(x_{t})) $
 for both state and space variable in the ghost and anti-ghost fields.

Pair trading is a root of a superstatistical arbitrage theory in which have a deep connection with the new model of DSGE in the superstatistic theory. The source of the financial market equilibrium comes from a pair of similar stock $(x_{t},y_{t})$ which is based on a pattern of mean-reversion or OU-equation or Fokker-Planck equation in the statistical physics.

\subsection{Moduli state space model in -14 Dimensions Superstatistical Theory}

We separate the system of the financial market into 2 parts, the first is the state part $X_{t}$ of the hidden demand state and the second is a the space part $Y_{t}$ of the observation of the supply space of the state space model. In this work, we use an algebraic equation from the algebraic topology and the differential geometry as a main tool for the definition of a new mathematical object for an arbitrary opportunity in the DSGE system of macroeconomics.

\begin{figure}[!t]
\centering
\includegraphics[width=0.48\textwidth]{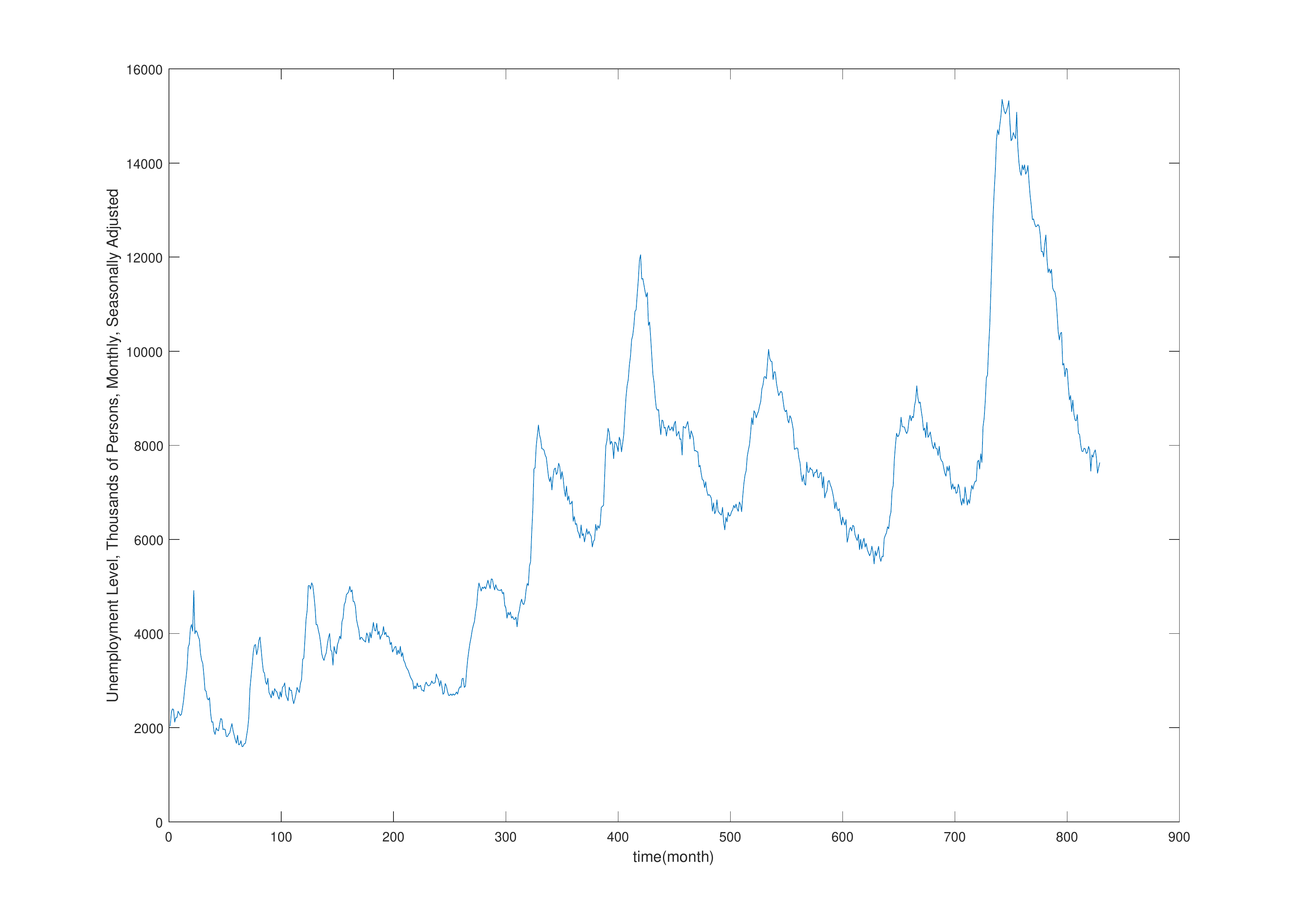}
\includegraphics[width=0.48\textwidth]{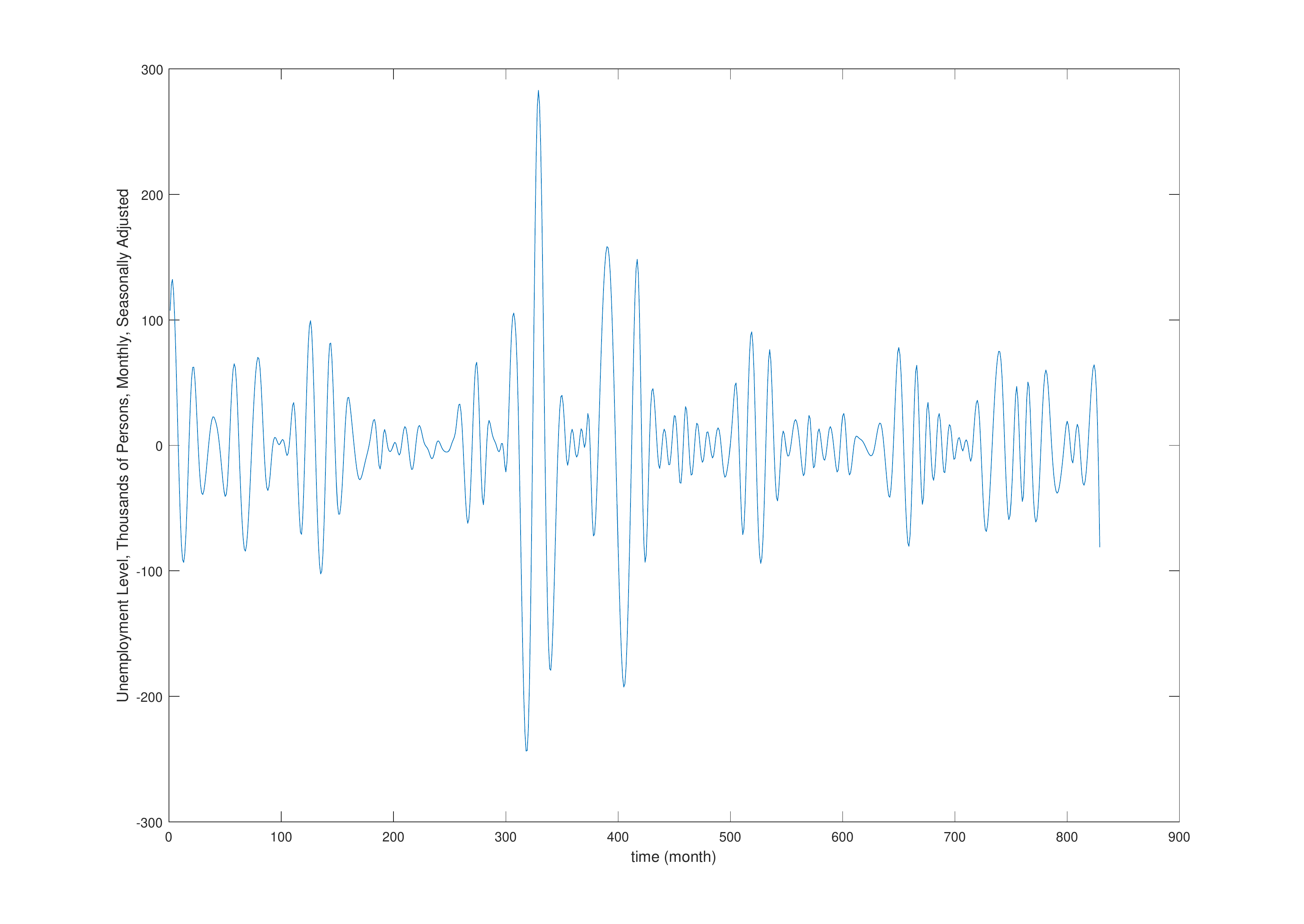}
\caption{Picture shows on the left panel the unemployment rate of the US-macroeconomy from 1945 to 2017, monthly data. We can notice a business cycle from a time series data of the unemployment rate. On the right panel, we use the moduli state space model of $(IMF-ITD)-chain-(1,4)$-transformation \cite{cohomo} to detect a business cocycle.}
\label{fig9}
\end{figure}

\begin{Theorem}
When market is in equilibrium , we have
\begin{equation}
s^{2}=0\,\leftrightarrow\,H^{-14}(\mathcal{A},s)=0.
\end{equation}
\end{Theorem}
{\em Proof: }

We are in the hidden 14 dimensional modeling in which each element of the cohomology group cancels each other when the market in equilibrium. In \cite{bv}, we have an element of the BV-cohomology group as a superprobability of the transition non-oriented superstate of the ghost fields, $f,g$ in financial market as

\begin{equation}
H^{-14}(\mathcal{A},s)=\{ \int \alpha_{k}(\Phi_{i}^{+}(y_{t}))+\beta_{k}(\Phi_{j}(x_{t}))| \Phi_{i}^{+}(y_{t}), \Phi_{j}(x_{t})\in \mathcal{O}\subset \mathcal{A} \},
\end{equation}
where $\alpha_{k},\beta_{k}$ are cocycles of the ghost and anti-ghost fields $\Phi_{j}(x_{t}),\Phi_{i}^{+}(y_{t})) \in \mathcal{A}.$ We have a master equation in the financial suppermanifold $\mathcal{A}$ as

\begin{equation}
\int_{H^{-14}(\mathcal{A},s)} S(\Phi_{1}, \Phi_{2}, \cdots ,\Phi_{14})dt, \{\int Sdt,\int Sdt\}=0.
\end{equation}

Let $\mathcal{A}$ be a supermanifold of the financial market with the master equation $(\mathcal{A},s)$.
The associate invariant algebraic group property of the market with higher dimensions influencing the factor of the market can be expressed by using the BV- cohomology in the algebraic equation

\begin{equation}
H^{-k} (\mathcal{A},s) =0
\end{equation}

We divide the market into 2 separated sheets of the D-brane and the anti-D-brane of the embedded indifference curve of supply and the utility curve of demand. The interaction of 2 D-branes is induced from the trade between the supply and the demand as the general equilibrium point. We define the D-brane sheet of the market in the real dimension and the anti-self-duality (AdS) of the D-brane to the anti-D-brane is induced from the duality map from the supply to the demand.

 Let IS-LM be written by
\begin{equation}
  y_{t} =\alpha_{t} +\beta_{t} x_{t} +  \epsilon_{t}
\end{equation}

Take a ghost  functor $\Phi_{i},\Phi_{i}^{+}:  X\rightarrow (\mathcal{A},s)\simeq [X,S^{\pm k}]$,

\begin{equation}
  \Phi_{i}( (y_{t} - \alpha_{t})- \beta_{t} x_{t} \simeq  \epsilon_{t}).
\end{equation}

Let $\epsilon_{t}$ be a real present shock from economics and $\epsilon_{t}^{\ast}$ be the expected shock in the future. In the equilibrium, we assume steady stead of the macroeconomics with no shock, i.e. $\epsilon_{t}^{2}=0$. Let 0 be a space of the equilibrium and let it be equivalent to the moduli state space model of the supply space $Y_{t}$ and the demand space $X_{t}$ of the market with $\epsilon_{t}^{2}=<\epsilon_{t},\epsilon_{t}^{\ast}>\simeq Y_{t}/X_{t}$. Consider a short exact sequence of the macroeconomics in the general equilibrium with the market risk $\beta_{t}$ of the sudden shock in the demand side and transfer it into the supply side of the economics and let systematics risk $\alpha_{t}$ be a shock on both sides with price sticky market. The shock in economics induces a business cycle which we can notice from the empirical data of the US unemployment from 1945 to present(see Fig. \ref{fig9}.)

\begin{equation}
0\longrightarrow \mathbb{Z}/2 \longrightarrow X_{t} \stackrel{\beta_{t}}{\longrightarrow} Y_{t}  \stackrel{\alpha_{t}}{\longrightarrow} Y_{t}/X_{t}
\longrightarrow 0,
\end{equation}

where we use the well-known exact sequence of the spinor

\begin{equation}
0\longrightarrow \mathbb{Z}/2 \longrightarrow Spin(n):=X_{t} \stackrel{\beta_{t}}{\longrightarrow} SO(n):=Y_{t}  \stackrel{\alpha_{t}}{\longrightarrow} Y_{t}/X_{t}
\longrightarrow 0.
\end{equation}
From this comparison between 2 exact sequences, we get the moduli state space model $X_{t}$ as a behavior of the trader on the demand side defined by the spin group and the Pauli matrix. The observation space of the supply side of the market is a physiology space of time series data which is spanned by basis of the equivalent class of spin, $<<[s_{1}],[s_{2}],[s_{3}],[s_{4}]>>:=Y_{t}$. In the macroeconomics model of the market, this space can be defined by the supply side of the market with the Lie group  $Y_{t}:=SO(n)=Spin(n)/\mathbb{Z}_{2}$. The moduli group $\mathbb{Z}_{2}$ defines a state up and down of the underlying financial time series data. When the market is in the equilibrium, the short exact sequence will induce an infinite exact sequence of the market cocycles $\beta_{t}$ and $\alpha_{t}$.

Now, we define an expected ghost field in the superspace of time series data and the superpoint in time series data of the underlying financial time series data.

\begin{Definition}
Let $S^{-1}$ be a space of unit circle with hidden 1-dimension. The minus sign means that this space is a future space and not a real space. It is induced from the expectation field underlying one factor as one dimension of the market factor. We explicitly define an associate real number with negative rank $\mathbb{R}^{-n}$ and a complex number with the negative dimension $\mathbb{C}^{-n}$ and we define $S^{-1}\simeq {S^{1}}^{\ast}$ , where $\ast$ is the mirror symmetry operation in the supersymmetry. The algebraic operation is the same with the positive dimension but the quantity of data underlying that space is not real but induced from the forecasting system or from the expectation of the traders in the financial market. When we come to present time, all negative dimensions will interact with positive real dimensions and fuse to the superpoint in the superspace in time series data.
\end{Definition}

\begin{Definition}
Let $S^{0}$ be a superpoint in superspace underlying time series data of the moduli state space model $(x_{t},y_{t})$. We define $S^{0}$ by the interaction between the unit cycle $S^{1}$ of the space of the present value of time series data and $S^{-1}$ of the future expectation value of the superspace of time series data.
\begin{equation}
S^{0}=S^{-1} \vee S^{1}=S^{-2}\vee S^{2} \cdots S^{-k} \vee S^{k}
\end{equation}

\end{Definition}

\begin{Definition}
Let $S^{-k}$ be a superspace underlying time series data of the moduli state space model $(x_{t},y_{t})$. Let $\mathcal{A}$ be a superspace of time series data with its ground field of the master equation $s=\{\int Sdt,-\}$ We define a BV cohomology group for the superspace in time series data a with negative $k$ dimension $H^{-k}(\mathcal{A},s)$ by using a homotopy class as a functor from $TOP$ to $GROUP$
\begin{equation}
 H^{-k}(\mathcal{A},s)=[(\mathcal{A},s),S^{-k}].
\end{equation}
The meaning of the cohomology group of the negative dimension is the superdistribution of the underlying superspace in time series data as the probability superdistribution of the future events.
\end{Definition}

We define the superspace of the financial market as the hidden layer of the extra dimensions of the Kolmogorov space in time series data. By this definition, given short exact sequence of the moduli state space model above, we induce an associate short exact sequence in the superspace layer of the sheave cohomology,

 \begin{equation}
\begin{CD}
0 @>>> (\mathcal{A},s)   @>>>S^{14}    @>>>S^{11}  @>>>S^{11}/S^{14}\sim S^{11}\vee S^{-14}\sim S^{-3}   @>>j>0    \sim \epsilon_{t}^{\ast} \\
@VVV @VV \Phi_{i} V @VVV   @VVV   @VVS^{3}\vee S^{-3}\sim S^{0}V  @VV  <\epsilon_{t},\epsilon_{t}^{\ast}>V\\
0 @>>>    \mathbb{Z}/2 @>>> X_{t}@>\beta_{t}>> Y_{t} @> \alpha_{t}>> Y_{t}/X_{t}\sim S^{3}@>>i> 0 \sim \epsilon_{t}
\end{CD}
 \end{equation}

where $i,j$ are the projection maps from the expected observation superspace $S^{\pm 3}$ of spinor field in time series data to shock and expected shock in time series data; when the equilibrium holds, i.e. $\epsilon_{t}\sim 0$,

\begin{equation}
j: S^{-3} \rightarrow \epsilon_{t}^{\ast} \hspace{1cm},j: S^{3} \rightarrow \epsilon_{t},
 \end{equation}

\begin{equation}
 <i,j>: S^{3}\vee S^{-3}\sim S^{0}\longrightarrow \epsilon_{t}^{2}.
 \end{equation}

In order to compute the market on the level of the invariance of the dimension of the cohomology of the sphere, we use the fact from the sheave cohomology that a short exact sequence induces an infinitely long exact sequence. The meaning of this infinite sequence comes from the homotopy path of the embedded point to the sphere and from the embedded sphere to torus and so on. We consider a short exact sequence of the moduli state space,

\begin{equation}
0 \rightarrow     \mathbb{Z}/2 \rightarrow  X_{t} \rightarrow Y_{t} \rightarrow Y_{t}/X_{t}\rightarrow 0
\end{equation}

This sequence induces an infinite sequence

\begin{equation}
0 \rightarrow     \mathbb{Z}/2 \rightarrow  X_{t} \rightarrow Y_{t} \rightarrow Y_{t}/X_{t}\rightarrow  H^{1}(\mathbb{Z}/2;\mathbb{Z}/2) \rightarrow  H^{1}(X_{t};\mathbb{Z}/2) \rightarrow  H^{1}(Y_{t};\mathbb{Z}/2) \rightarrow H^{1}(Y_{t}/X_{t};\mathbb{Z}/2)\rightarrow \nonumber
\end{equation}

\begin{equation}
 \rightarrow  H^{2}(\mathbb{Z}/2;\mathbb{Z}/2) \rightarrow  H^{2}(X_{t};\mathbb{Z}/2) \rightarrow  H^{2}(Y_{t};\mathbb{Z}/2) \rightarrow H^{2}(Y_{t}/X_{t};\mathbb{Z}/2)\rightarrow \nonumber
\end{equation}

\begin{equation}
 \rightarrow  H^{3}(\mathbb{Z}/2;\mathbb{Z}/2) \rightarrow  H^{3}(X_{t};\mathbb{Z}/2) \rightarrow  H^{3}(Y_{t};\mathbb{Z}/2) \rightarrow H^{3}(Y_{t}/X_{t};\mathbb{Z}/2)\rightarrow \nonumber
\end{equation}

\begin{equation}
 \rightarrow    \cdots \rightarrow  \nonumber
 \end{equation}
\begin{equation}
 \rightarrow  H^{11}(\mathbb{Z}/2;\mathbb{Z}/2) \rightarrow  H^{11}(X_{t};\mathbb{Z}/2) \rightarrow  H^{11}(Y_{t};\mathbb{Z}/2) \rightarrow H^{11}(Y_{t}/X_{t};\mathbb{Z}/2)\rightarrow \cdots
\end{equation}

In the hidden layer of the superspace in time series data, we induce an infinitely long exact sequence in the BV-cohomology of a negative dimension

\begin{equation}
(\mathcal{A},s)\rightarrow H^{-1}(\mathcal{A},s) \rightarrow H^{-2}(\mathcal{A},s)\longrightarrow \cdots \rightarrow H^{-14}(\mathcal{A},s)\longrightarrow \cdots .
\end{equation}
We also have an exact sequence
\begin{equation}
0 \rightarrow  (A,s)  \rightarrow S^{14}    \rightarrow S^{11}  \rightarrow S^{11}/S^{14}\sim S^{11}\vee S^{-14}\sim S^{-3}   \rightarrow 0
\end{equation}
in which we induce an infinite exact sequence of the cohomology of a negative dimension of a sphere,
\begin{equation}
0 \rightarrow  H^{-1}((A,s))  \rightarrow H^{-1}(S^{14})    \rightarrow H^{-1}(S^{11})  \rightarrow H^{-1}( S^{-3})   \rightarrow    H^{-2}((A,s))  \rightarrow H^{-2}(S^{14})    \rightarrow H^{-2}(S^{11})  \rightarrow\cdots
\end{equation}
When the market is in the general equilibrium, we have $H^{-k}(A,s)=0$ for all $k>0$. For the detail of the proof see \cite{bv}.

  The BV-cohomology group is

\begin{equation}
 \cdots \longrightarrow H^{-7}(\mathcal{A},s)
  \longrightarrow H^{-8}(\mathcal{A},s)
\longrightarrow  \longrightarrow H^{-9}(\mathcal{A},s)
\cdots
\longrightarrow H^{-14}(\mathcal{A},s)
\end{equation}
for the pullback functor of the effect of the Calabi-Yau  manifold.

Let $x_{t}\in X_{t}$ be a moduli state space of a D-brane and Anti-D-brane. Let $y_{t} \in Y_{t}$ be
 an observation space of the financial market.

\begin{figure}[!t]
\centering
\includegraphics[width=0.48\textwidth]{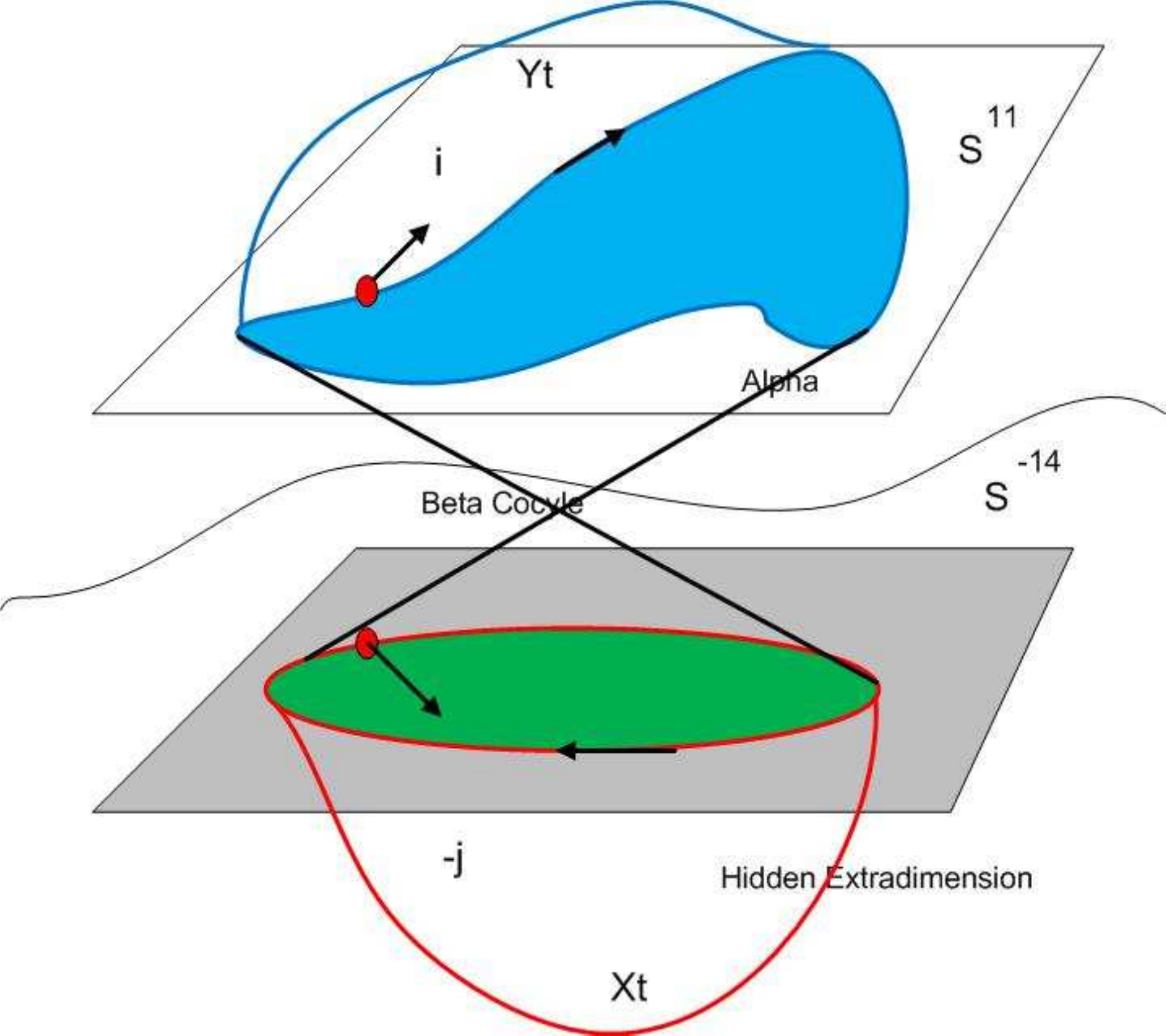}
\includegraphics[width=0.48\textwidth]{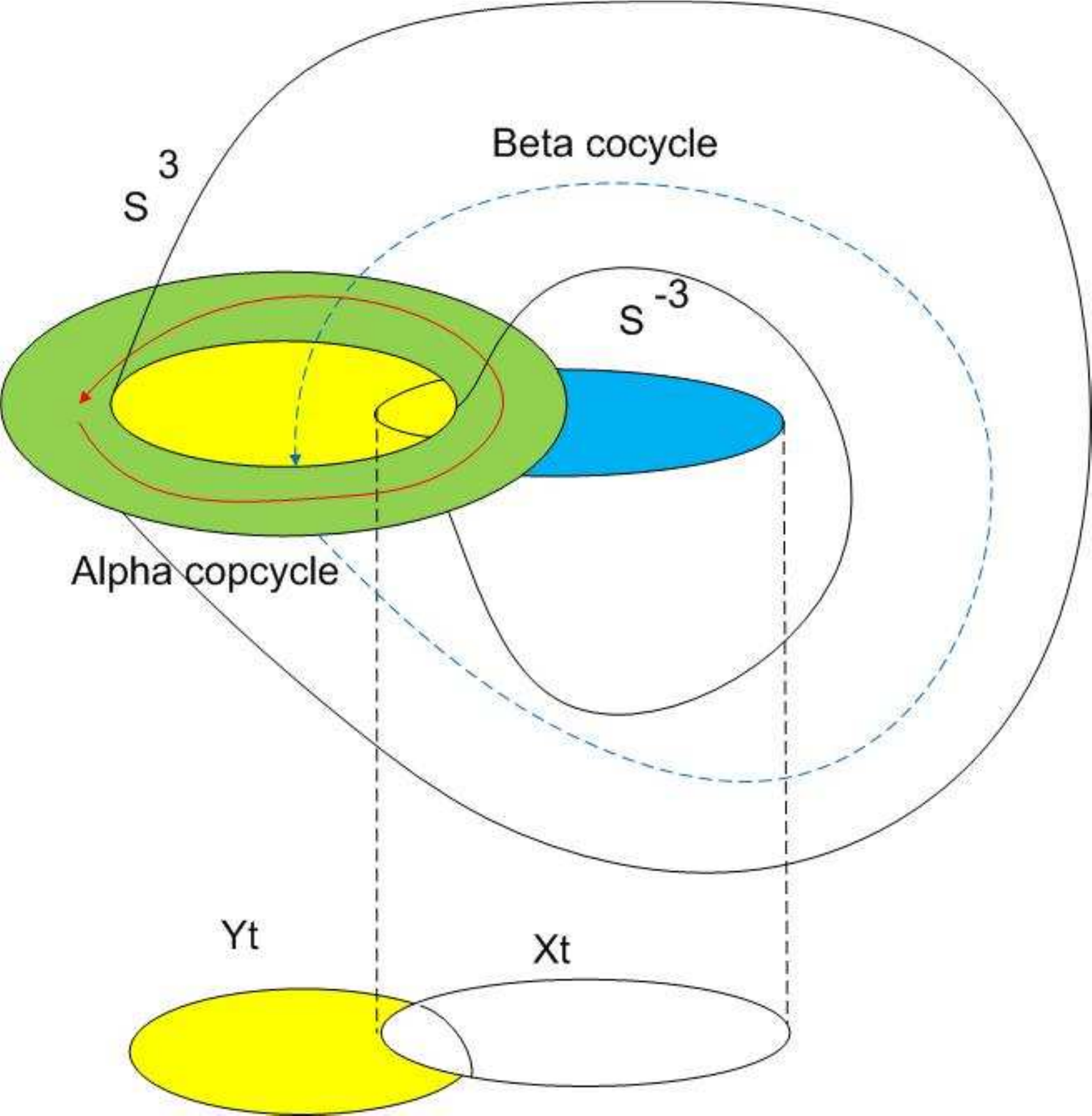}
\caption{Picture shows on the left panel 2 layers of the market. The upper layer is a
four dimensional model of the financial market $S^{3}$ as torus $S^{1}\times S^{1}$ with the Hopf fibration.
The lower layer is a hidden layer of the market as a superspace in time series data. This layer is associated with an expectation layer of the market in which the induction from the interaction between 14-ghost fields of the behavior of the traders in the financial market on the expected 14-market factors influence the market future equilibrium point in the physiology of time series data. In this proof we need to define a spectral sequence of the space - so-called negative dimension unit sphere $S^{-1}$ as a space of the expectation ghost field in the financial market. In the economic modeling of the supermanifold, there is not yet defined the negative dimension $S^{-1}$ because no one starts to use the supersymmetry theory and the BV-cohomology in the economic yet. On the right panel, where we use the BV-cohomology group of the negative order 3 to compute the space of $X_{t}$, the general equilibrium of the market exists as knots of the market risk $\beta_{t}$ and the systematics risk $\alpha_{t}$. We induce the general equilibrium point of the market as the superpoint in the superspace of the financial market with $S^{0}\sim S^{-3}\vee S^{3}.$ In this model, we can generally associate the hidden sphere with any dimension up to the market factors influencing the underlying financial market by using the sheave cohomology as the main tool for the computing of the market equilibrium.}
\label{fig8}
 \end{figure}

\begin{equation}
 H^{-14}(x_{t}  \sim y_{t}) =0=\alpha_{t}y_{t}-\beta_{t}x_{t}:=\sum_{i=1}^{2}n_{i}g_{i}:=pp(\epsilon^{2}_{t}):=
 <\alpha_{t},\beta_{t}><x_{t},y_{t}>^{\ast},
\end{equation}
where $\alpha_{t}y_{t}$ is a group operation of the reversed direction of the translation, i.e.
 $\alpha_{t} y_{t}:=g(y_{t})=y_{t}-\alpha_{t} $ and $\beta_{t}x_{t}$ is a group action of the spinor rotational group. The equilibrium is the anomaly cancelation and the net area will be contractible to the point.
The curvature in the area of the market in the equilibrium state is canceled to each other and the net sum is zero.
The meaning of the zeros in the BV-cohomology group is the analogy with the zero cohomology group of a plane without an obstruction component. We explicitly define a ghost field of the financial market in the loop space of time series data (see Fig. \ref{fig8}) by using a fundamental group over a homotopy path
\begin{equation}
\Phi_{i}: Y_{t}\times I \rightarrow S^{11}, \Phi_{i}^{+}: X_{t} \times I\rightarrow S^{-14}
\end{equation}
where $S^{-1}$ is a unit sphere with a negative dimension(non-oriented supermanifold of its parity of $S^{1}$) mod 2.

We explicitly define an element of a ghost field by using a group action over a cotangent bundle of the supermanifold as a cocycle $\alpha_{t}$ and a business cycle $\beta_{t}$ .

We obtain the result that
\begin{equation}
Y_{t}/X_{t}=\frac{\alpha_{t}[y_{t}]}{\beta_{t}([x_{t}])}\simeq [{\epsilon_{t}}^{\ast}] \in S^{-3}.
\end{equation}
We explicitly define 14 ghost and 11 anti-ghost fields in the financial market with the demand side of the state space with $x_{t}  \in H^{-14}(X_{t}):=[X_{t},S^{-14}]$ and with the supply side $y_{t}\in H^{11}(Y_{t})=[Y_{t},S^{11}]$. In the market equilibrium, we have
\begin{equation}
S^{3}|_{\epsilon_{t}} \vee S^{11}\vee S^{-14} |_{\epsilon_{t}^{\ast}}\sim S^{3}\vee S^{-3} \simeq S^{0}\ni \epsilon_{t}^{2}=0.
\end{equation}
If the market is in the non-equilibrium, then there exists a cocycle $(\alpha_{t},\beta_{t})$ and $\epsilon_{t}\neq 0$ such that

\begin{equation}
y_{t}=\alpha_{t}+\beta_{t}x_{t}+\epsilon_{t}.
\end{equation}

\section{Discussion and Conclusion}
In this work, we provided an explanation of an asymmetric model of the financial market and its duality. We explained  why the financial market equilibrium will give zeros of the BV-cohomology group of negative 14 dimensions. We use a new G-theory to explain the financial market in 14 dimensions. Our model is not the as the $E_{8}\times E_{8}$ and $SO(32)$ model of the financial market over 11 dimensions, M-theory. We assume that under the record of financial time series data, there exists a superspace as a supermanifold of the ghost and anti-ghost fields with the parity mode $2$ induced from the supply and demand ghost fields of the market. This fields wrap to each other like the D-brane interaction with the anti-D-brane model in the unified theory. We study the ghost and anti-ghost fields in the superspace of the market in the mirror symmetry with the anomaly as an arbitrary opportunity. We define a new BV-cohomology for the financial market with its market supersymmetry and cohomology as its duality between the supply and demand as a superspace of time series data. We use the supermanifold modeling in the space of time series data when the space is in the non-oriented hidden ghost field. We define a new equation for the financial market in which one can unify the microeconomic and macroeconomic theories by using the Chern-Simon and Yang-Mills theories with the BV-cohomology. We use the approach of the G-theory in the unified theory in 14-dimensional model of the birth of the universe to unify the economic parameters in the superspace of time series data with the transition superprobability for the quantized ghost field in the anomaly Chern-Simon current induced from the coupling between the market cocycle of the behavior of the traders in the financial market. We write a Yang-Mills-Chern-Simon-master equation in general from the superlagrangian with the least action principle. We use the BV-cohomogy to prove the existence of the dynamical stochastic general equilibrium point in the financial market when the cohomology is in the exact sequence and the hidden 14 dimensions are vanishing with arbitrary opportunity existing as a curvature of an embedded superspace of time series data in the form of the Calabi-Yau space in time series data in the reversed direction of the time scale. In future work, it is possible to follow the model and use the computer simulation to visualize the geometry of the market in the equilibrium and do some empirical analysis over major 14 hidden factors in the supply and 11 factors in the demand side according to our high dimensional economic modeling. In order to find a market general equilibrium as arbitrary opportunity, we use the connection and the covariant derivative to error of the forecasting market of all the hidden factors and take it to zeros according to our providing theorem of the 14-th BV cohomology group. We conclude that the vanishing of the cohomology group will satisfy the stability of the topological property of the market and the empirical analysis will show us the precise system of the complicated equations for the prediction of all the parameters from all the economic factors in which one can influence an equilibrium point of the market. We got a result from our research that the market equilibrium will produce a spinor field of the systematic risk and the market risk as the coupling market cocycle and vanishing as an induced spinor field from the general equilibrium point of the financial market in the form of an arbitrary opportunity over the market mirror symmetric property. We exactly need 14 dimensions for the real economic modeling of the global financial market and for the real computer simulation results on the real data to visualize the general equilibrium model as our future work in order to verify our theoretical modeling of the cohomology group of the superspace of the market.

\section*{Acknowledgment}
S. Capozziello acknowledges financial support of  INFN (iniziative specifiche TEONGRAV and QGSKY).
K. Kanjamapornkul is supported by the scholarship from the 100th Anniversary Chulalongkorn University Fund for Doctoral Scholarship.  The work was partly supported by VEGA Grant No. 2/0009/16. R. Pincak would like to thank the TH division at CERN for hospitality. We are particularly grateful to Erik Bartoš for the important comments and observations.\addcontentsline{toc}{section}{Acknowledgment}


\end{document}